\documentclass[12pt,preprint]{aastex}
\begin{document}

\title{Frequency spectrum of toroidal Alfv\'en mode in a neutron star
with Ferraro's form of nonhomogeneous poloidal magnetic field}

\author{
S.I. Bastrukov\altaffilmark{1,3}, H.-K. Chang\altaffilmark{1,2}, I.V. Molodtsova\altaffilmark{3},
E.-H. Wu\altaffilmark{1},
G.-T. Chen\altaffilmark{2}, S.-H. Lan\altaffilmark{1}}

\altaffiltext{1}{Institute of Astronomy,
  National Tsing Hua University, Hsinchu, 30013, Taiwan}

\altaffiltext{2}{Department of Physics,
  National Tsing Hua University, Hsinchu, 30013, Taiwan}

\altaffiltext{3}{Joint Institute for Nuclear Research, 141980 Dubna, Russia}

\begin{abstract}
 Using the energy variational method of magneto-solid-mechanical theory of a perfectly conducting elastic medium threaded by magnetic field, the frequency spectrum of Lorentz-force-driven global torsional nodeless vibrations of a neutron star with Ferraro's form of axisymmetric  poloidal nonhomogeneous internal and dipole-like
 external magnetic field is obtained and compared with that for this toroidal Alfv\'en mode in
 a neutron star with homogeneous internal and dipolar external magnetic field.
 The relevance of considered asteroseismic models to quasi-periodic oscillations of the X-ray flux during the ultra powerful outbursts of SGR 1806-20 and SGR 1900+14 is discussed.
\end{abstract}


\section{Introduction}
 In the context of recent discovery of quasi-periodic oscillations (QPOs) in the X-ray luminosity during the giant flare of SGR 1806-20 and SGR 1900+14 that were interpreted as being produced by torsional vibrations of quaking magnetars (Israel et al. 2005, Watts \& Strohmayer 2006), in (Bastrukov et al. 2009a, 2009b) several scenarios of the post-quake vibrational relaxation of a neutron star model with uniform internal and dipolar external magnetic field
  \begin{eqnarray}
  \label{e1.1}
   && B_r=B\cos\theta,\quad B_\theta=-B\sin\theta,\quad B_\phi=0,\quad r\leq R\\
   \label{e1.2}
  && B_r=B\left(\frac{R}{r}\right)^3\cos\theta,\quad B_\theta=-\frac{B}{2}\left(\frac{R}{r}\right)^3\sin\theta,
 \quad B_\phi=0,\quad r > R
 \end{eqnarray}
 have been studied on the basis of equations of Newtonian magneto-solid-mechanics
 \begin{eqnarray}
  \label{e1.3}
 && \rho{\ddot {\bf u}}=\frac{1}{c}[\delta {\bf j}\times {\bf B}],\quad
 \delta {\bf j}=\frac{c}{4\pi}[\nabla\times \delta {\bf B}], \quad \delta {\bf B}({\bf r},t)=\nabla\times [{\bf u}\times {\bf B}].
  \end{eqnarray}
   These equations describe the Lorentz-force-driven non-compressional ($\delta \rho=-\rho \nabla\cdot  {\bf u}=0$)
   fluctuations of star matter about axis of above fossil magnetic field ${\bf B}$
   in terms of fluctuating material displacements $\bf u$ (the basic dynamical variable of solid mechanics) and the magnetic field  $\delta {\bf B}$. It is implied that elastic stellar material is of an extremely high electrical conductivity\footnote{The decay time of equilibrium magnetic field of the neutron stars is much longer than the time intervals between X-ray bursts and periods of their quiescent pulsed emission (e.g. Bhattacharya \& van den Heuvel 1991, Chanmugham 1992, Goldreich and Reisenegger 1992), so
   that adopted approximation of an infinite electrical conductivity of the star matter is amply justified.}.
   The chief argument for interpreting QPOs during the outbursts of above mentioned SGRs (detected on descending branch of the giant flare light-curves) as caused by torsional vibrations of a solid star driven by restoring force of magnetic field stresses is that it is an ultra strong magnetic field frozen-in the entire volume of magnetars serves as the main energy source and promoter of their X-ray bursting seismic activity (e.g., Woods \& Thompson 2006, Mereghetti 2008) and, also, bearing in mind the fact that the very notion of torsional vibrations
   has come into theoretical seismology from the solid-mechanical theory of shear vibrations of an elastic sphere
   (e.g. Lapwood, \& Usami 1981,  Lay \& Wallace 1995, Aki \& Richards 2002, Stein \& Wyssesson 2003).
   It worthy noting that theoretical investigations of non-radial torsional Alfv\'en oscillations of a fluid star in its own homogeneous magnetic field have a long story that was started, to the best of our knowledge, in
   works of Jensen (1955) and Plumpton \& Ferraro (1955). Remarkably that in the latter paper, by emphasizing the
   basic discovery of Alfv\'en that perfectly conducting fluid threaded by magnetic field behaves like anisotropic elastic medium capable of transmitting mechanical disturbance by transverse hydromagnetic waves (e.g. F\"althammar 2007), it is argued that eigenfrequency problem of such vibrations can be tackled on the basis of equation
  \begin{eqnarray}
  \label{e1.4}
  && \rho{\ddot {\bf u}}=\frac{1}{4\pi}[\nabla\times[\nabla\times [{\bf u}\times {\bf B}]]]\times {\bf B},\quad \nabla\cdot  {\bf u}=0,\quad \nabla\cdot {\bf B}=0
  \end{eqnarray}
  which, as is evident, follows form equations (\ref{e1.3}). It is clear from this last equation that the frequencies
  of Alfv\'en oscillations must substantially depend on both geometrical configuration of internal
  equilibrium magnetic field ${\bf B}$ and analytic form of oscillating field of material displacements ${\bf u}$.
  With this obvious observation in mind, in (Bastrukov et al. 2009a, 2009b) focus
  was laid on non-investigated before regime of large lengthscale nodeless Alfv\'en oscillations, both global (in the entire spherical volume of star) and  crustal (locked in the peripheral finite-depth spherical layer).
  The most conspicuous feature of this regime is that the radial dependence of oscillating material displacements field ${\bf u}$ has no nodes. In a star undergoing global nodeless torsional oscillations, which are of particular interest for our present discussion, the fluctuating material displacements are described by the toroidal field of the form (Bastrukov et al 2007a, 2007b, 2009a)
  \begin{eqnarray}
  \label{e1.5}
  && {\bf u}({\bf r},t)=[\mbox{\boldmath $\phi$}({\bf r})\times {\bf r}]\,\alpha(t),\quad \alpha(t)=\alpha_0\cos\omega t,\\
  \label{e1.6}
  && \delta {\bf v}({\bf r},t)={\dot {\bf u}}({\bf r},t)=[\delta \mbox{\boldmath $\omega$}({\bf r},t)\times {\bf r}], \quad
  \delta\mbox{\boldmath $\omega$}({\bf r},t)=\mbox{\boldmath $\phi$}({\bf r})\,{\dot \alpha}(t),\\
 \label{e1.7}
 && \mbox{\boldmath $\phi$}({\bf r})=\nabla\chi({\bf r}),\quad \nabla^2\chi({\bf r})=0,\quad \chi({\bf r})=
 \frac{{\cal A}_\ell}{\ell+1}\,r^\ell\,P_\ell(\cos\theta)
 \end{eqnarray}
 where $P_\ell$ is the Legender polynomial of degree $\ell$ and the other symbols have their usual meaning.
 Fig.1 illustrates the nodeless character of displacements in the star undergoing global
 non-radial differentially rotational, torsional, shear vibrations about polar axis in quadrupole  and octupole overtones.

 \begin{figure}[h]
\centering\
\includegraphics[width=9.0cm]{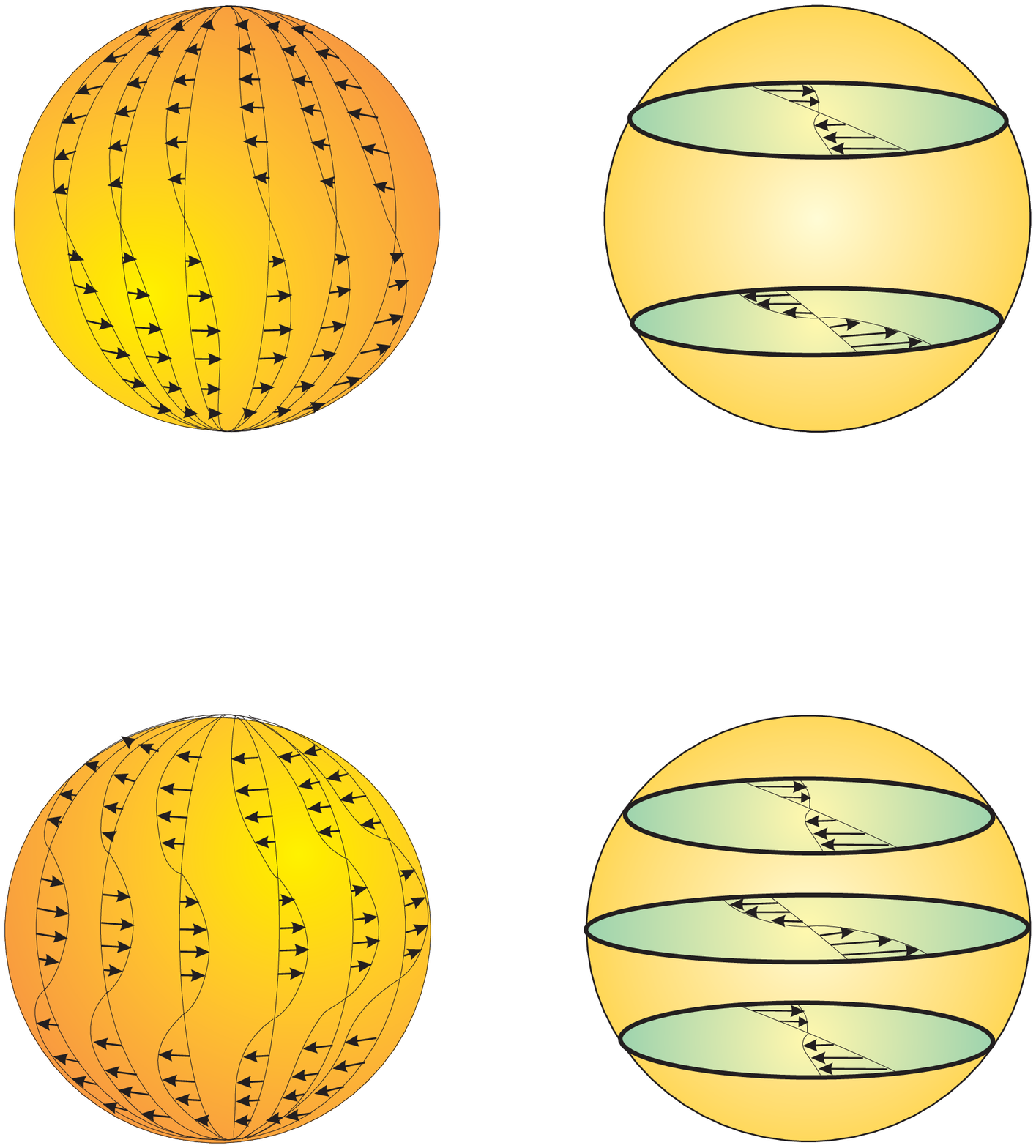}
\caption{\small
 Material displacements in the neutron star undergoing axisymmetric global torsional nodeless  vibrations in quadrupole $\ell=2$ and octupole $\ell=3$ overtones.}
\end{figure}

 With the aid of the Rayleigh's energy method which is expounded below it was found that discrete
 frequencies of such vibrations is given by the following one-parametric spectral formula (Bastrukov et al. 2009a)
 \begin{eqnarray}
 \label{e1.8}
 && \omega(_0a^t_\ell)=\omega_A\left[(\ell^2-1)\frac{2\ell+3}{2\ell-1}\right]^{1/2},\,\,
 \omega_A=\frac{v_A}{R}=\sqrt{\frac{RB^2}{3M}},\\
 \label{e1.9}
 && \quad v_A=\frac{B}{\sqrt 4\pi \rho},\quad M=\frac{4\pi}{3}\rho\,R^3
 \end{eqnarray}
 the only parameter of which is the Alfv\'en frequency, $\omega_A$, of shear magneto-elastic oscillations of perfectly conducting stellar matter matter pervaded by magnetic field $B$ in the star of radius $R$ and mass M.
 It must be emphasized, however, that this theoretical
 spectrum does not properly match the QPOs in the X-ray flux from flaring SGR 1806-20 and SGR 1900+14.
 One of reasons of this discrepancy may be inadequate assumption about homogeneous configuration of internal magnetic field and perhaps the most efficient way to clarify this conjecture is to investigate a model with geometrically
 different configuration of axisymmetric internal magnetic field.
 Before so doing it seems worth noting that the model of a star with {\it homogeneous} internal and dipolar external magnetic field has come into focus in astrophysics after seminal work of Chandrasekhar and Fermi (1953) in which the effect of mechanical flattening of the star at the poles of such magnetic field has been disclosed. Shortly after, similar conclusion has been drawn in outstanding paper of Ferraro (1954), but on the basis of star model with substantially {\it nonhomogeneous} internal and dipole-like
 axisymmetric poloidal magnetic field
  \begin{eqnarray}
  \label{e1.10}
   && B_r=\frac{1}{r^2\sin\theta}\frac{\partial U}{\partial \theta},\quad B_\theta=-\frac{1}{r\sin\theta}\frac{\partial U}{\partial r},\quad B_\phi=0,\\
   \label{e1.11}
   && U=U_{in}=\frac{B}{4R^2}\,r^2(3r^2-5R^2)\sin^2\theta,\quad r\leq R,\\
   \label{e1.12}
   && U=U_{ex}=\frac{B}{2r}\,R^3 \sin^2\theta,\quad r> R
  \end{eqnarray}
 where $B$ stands for the magnetic field intensity at the poles and $\nabla\cdot {\bf B}=0$ as should be the case\footnote{It may be noteworthy that
 magnetic energy stored
 in the star volume, $W=(1/8\pi)\int B^2\,d{\cal V}$, with this nonhomogeneous (nh) internal magnetic field $W_{nh}=(69/252)B^2R^3\approx 0.24B^2R^3$
 is somewhat larger than in the star with homogeneous (h) magnetic field  $W_h=(1/6)B^2R^3\approx 0.17B^2R^3$.}. The meridian cross section of Ferraro's model of the star is sketched in Fig.2.

 \begin{figure}[h]
\centering\
\includegraphics[width=9.0cm]{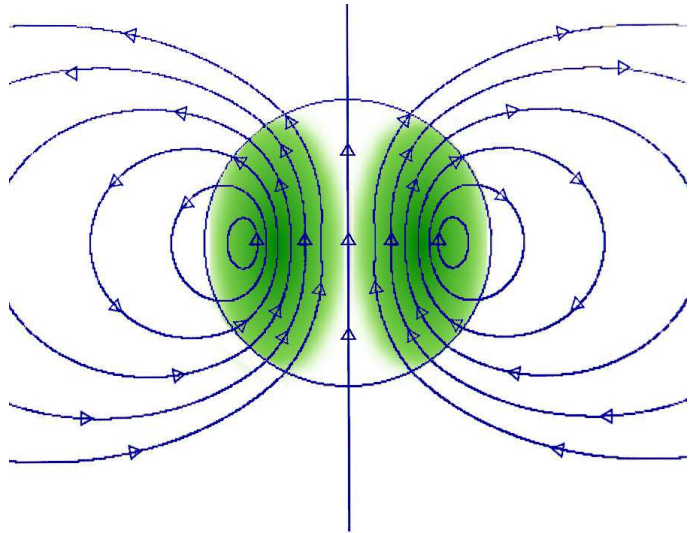}
\caption{\small
The meridional cross section of a neutron star with Ferraro's form of inhomogeneous poloidal internal and dipolar external magnetic field whose components are continuous on the star surface, contrary to a highly idealized
star model with homogeneous internal and dipolar external magnetic field.}
\end{figure}

 Over the years,
 the different aspects, both mathematical and astrophysical, of Ferraro's model have been the subject of extensive investigations (e.g. Chandrasekhar \& Prendergast 1955, Robetrs 1955, Chandrasekhar 1956, Mestel 1956, Ledoux \& Walraven 1958, Monaghan 1965, Ledoux \& Renson 1966, Sood \& Trehan 1970, Goossens 1972, Goossens, Smeyers \& Denis 1976).
 The effects of Ferraro's configuration of magnetic field (and
 magnetic fields of similar geometrical configuration) on the equilibrium shape, vibrational behavior and electromagnetic activity of pulsars and magnetars are considered in (Roberts 1981, Ioka 2001, Braithwaite \& Spruit 2006, Geppert \& Rheinhardt 2006, Haskell et al 2008,  Lee 2008; Broderick \& Narayan 2008, see also references therein).

 In this work we focus on the non-studied before regime of global torsional Alfv\'en nodeless
 vibrations of neutron star about axis of Ferraro's magnetic field (\ref{e1.10})-(\ref{e1.12}). In Section 2, the frequency spectrum of this toroidal mode is derived and compared with the frequency spectrum (\ref{e1.8}) of the neutron star model with homogeneous internal magnetic field. In Section 3, the obtained spectral formula for the frequency is analyzed numerically in juxtaposition with data on QPOs during
 the flare of SGR 1806-20 and SGR 1900+14. Section 4 briefly accounts for the net outcome of this work.
 Technical details of analytic computations can be found in Appendix.

\section{Global Alfv\'en torsional nodeless oscillations of neutron star in its own poloidal magnetic field of Ferraro's form}

 In the model under consideration a neutron star is identified with a finite spherical mass of an elastic solid, regarded as an incompressible continuous medium of uniform density $\rho$ and an infinite electrical conductivity, whose vibrations under the action of Lorentz magnetic force are governed by equations of magneto-solid-mechanics (\ref{e1.2}) which can conveniently be represented in the following equivalent tensor form (e.g. Mestel 1999)
  \begin{eqnarray}
  \label{e2.1}
  &&\rho{\ddot u}_i=\nabla_k\, \delta M_{ik},\quad \delta M_{ik} =\frac{1}{4\pi}[B_i\,\delta B_k+B_k\,\delta B_i-(B_j\,\delta B_j)\delta_{ik}], \\
  && \delta B_i({\bf r},t)=(B_k \nabla_k)u_i-(u_k\nabla_k) B_i,\quad \nabla_i u_i=0
  \end{eqnarray}
  where $\delta M_{ik}$ stands for the Maxwell's tensor of magnetic field stresses.
  The energy balance in the process of vibrations is controlled by equation
  \begin{eqnarray}
 \label{e2.2}
 && \frac{\partial }{\partial t}\int \frac{\rho{\dot u}^2}{2}\,d{\cal V}
 =- \int \delta M_{ik}\,{\dot u}_{ik}\,d{\cal V}=-\frac{1}{8\pi}\int [B_i\,\delta B_k+B_k\,\delta B_i]\,
 [\nabla_i {\dot u}_k + \nabla_k {\dot u}_i]\,d{\cal V},\\
 \label{e2.3}
 && {\dot u}^2={\dot u}_i\,{\dot u}_i,\quad {\dot u}_{ik}=\frac{1}{2}[\nabla_i {\dot u}_k+\nabla_k {\dot u}_i],\quad
 {\dot u}_{kk}=\nabla_k {\dot u}_k=0.
  \end{eqnarray}
  To compute the eigenfrequency of toroidal Alfv\'en mode in question we take advantage
  of the Rayleigh's energy method which has been utilized in our previous above mentioned investigations.
  The key idea of this method consists in separable representation of fluctuating variables such as
  the vector field of material displacements $u_i({\bf r},t)$ and the tensor field of shear strains $u_{ik}({\bf r},t)$
  \begin{eqnarray}
 \label{e2.4}
 u_i({\bf r},t)=a_i({\bf r})\,{\alpha}(t),\quad u_{ik}({\bf r},t)=a_{ik}({\bf r})\,\,{\alpha}(t),\quad a_{ik}({\bf r})=\frac{1}{2}[\nabla_i a_k({\bf r})+\nabla_k a_i({\bf r})].
 \end{eqnarray}
  With this form of $u_i$, the magnetic flux density $\delta B_i({\bf r},t)$ and the tensor field of fluctuating magnetic field stresses $\delta M_{ik}({\bf r},t)$ are represented in a similar manner
  \begin{eqnarray}
  \label{e2.5}
 && \delta B_i({\bf r},t)=b_i({\bf r})\,\alpha(t),\quad b_i({\bf r})=(B_k \nabla_k)a_i-(a_k\nabla_k) B_i,\\
 \label{e2.6}
 && \delta M_{ik}({\bf r},t)=[\tau_{ik}({\bf r})-\frac{1}{2}\tau_{jj}\delta_{ik}]\alpha(t),\quad \tau_{ik}({\bf r})=\frac{1}{4\pi}[B_i({\bf r})\,b_k({\bf r})+B_k({\bf r})\,b_i({\bf r})].
 \end{eqnarray}
 The gist of this multiplicative decomposition of fluctuating variables is that on substituting (\ref{e2.4})-(\ref{e2.6}) in (\ref{e2.2}) this latter equation is reduced to equation for time-dependent amplitude $\alpha(t)$ having the well-familiar form
 \begin{eqnarray}
\label{e2.7}
 && {\cal M}{\ddot \alpha}(t)+{\cal K}_m\alpha(t)=0,\quad {\cal M}=\int \rho\, a_i\,a_i d{\cal V},\\
 \label{e2.8}
 && {\cal K}_m=\int \tau_{ik}\,a_{ik}\,d{\cal V}=\frac{1}{8\pi}\int [B_i\,b_k+B_k\,b_i]\,[\nabla_i\,a_k+\nabla_k\,a_i]\,d{\cal V}.
  \end{eqnarray}
  Thus, from technical argument, the computation of frequency $\omega=[{\cal K}/{\cal M}]^{1/2}$ is reduced to
  calculation of integral parameters of inertia ${\cal M}$ and stiffness ${\cal K}_m$ with
  the toroidal field of instantaneous, time-independent, displacements
  \begin{eqnarray}
 \label{e2.9}
 && {\bf a}_t=A_t\,\nabla \times  [{\bf
 r}\,r^\ell\,P_\ell(\zeta)]:\,\, a_r=0,\,\,\, a_\theta=0,\,\,\, a_\phi=A_t r^{\ell}(1-\zeta^2)^{1/2}\frac{dP_\ell(\zeta)}{d\zeta}
 \end{eqnarray}
 and the magnetic field of Ferraro's form whose spherical components inside the star are
  \begin{eqnarray}
   \label{e2.10}
  && B_r=\frac{3B}{2R^2}\,r^2\left(r^2-\frac{5}{3}R^2\right)\cos\theta,\quad B_\theta=-\frac{3B}{2R^2}\,\left(2r^2-\frac{5}{3}R^2\right)\sin\theta,\quad \quad B_\phi=0.
  \end{eqnarray}

  \begin{figure}[h]
\centering\
\includegraphics[width=12.0cm]{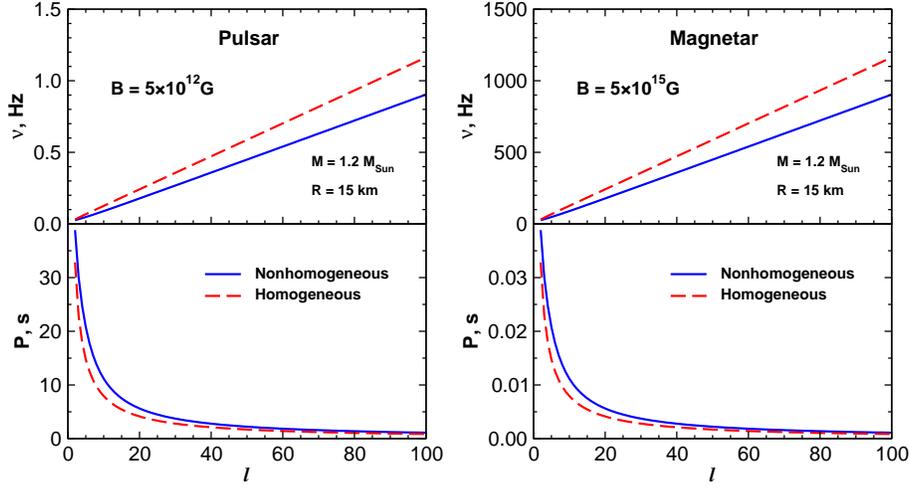}
 \vspace*{0.3cm}
 \caption{\small
 Frequency and period as functions of multipole degree $\ell$ of global torsional Alfv\'en vibrations
 of neutron stars with the Ferraro's shape of internal poloidal magnetic field.}
 \end{figure}

  The torsional inertia ${\cal M}$ as a function of multipole degree $\ell$ of nodeless differentially rotational vibrations in question is given by (Bastrukov et al, 2007, 2008)
\begin{eqnarray}
 \label{e2.11}
 {\cal M}=4\pi\rho A_t^2 R^{2\ell+3}\,m_\ell,\quad m_\ell=\frac{\ell(\ell+1)}{(2\ell+1)(2\ell+3)}.
 \end{eqnarray}
 To avoid destructing attention from basic inferences of this work, we place all technical
 details of tedious but simple computations of integrals for ${\cal K}_m$ in Appendix A. The
 final expression for this coefficient can be represented in the form
 \begin{eqnarray}
 \label{e2.12}
 {\cal K}_m=B^2\,A_t^2\,R^{2\ell+1}\, k_\ell,\quad
  k_\ell=\frac{\ell \left(\ell^2-1 \right) \left(5\ell^3+7\ell^2+59\ell+84 \right)   }
 {2\left(4\ell^2-1 \right)\left(2\ell+3 \right) \left(2\ell+5 \right) }.
 \end{eqnarray}
 And for the frequency spectrum of global nodeless torsional Alfv\'en vibrations nodeless of the neutron star with Ferraro's form of nonhomogeneous internal magnetic field we obtain
 \begin{eqnarray}
 \label{e2.14}
&& \nu(_0a^t_\ell) =\nu_A\left[\frac{\left(\ell-1 \right) \left( 5\ell^3+7\ell^2+59\ell+84\right) }{2\left(2\ell-1 \right) \left(2\ell+5 \right)}\right]^{1/2},\\
&& \nu=\frac{\omega}{2\pi},\quad \omega_A=\frac{v_A}{R},\quad v_A=\frac{B}{\sqrt{4\pi\rho}},\quad \omega_A=B\sqrt{\frac{R}{3M}},\quad M=\frac{4\pi}{3}\rho\,R^3.
\end{eqnarray}
 It follows that the lowest overtone of this toroidal Alfv\'en mode is of quadrupole degree, $\ell=2$. At $\ell=1$, the parameter of magneto-mechanical rigidity of neutron star matter cancels, ${\cal K}_m(_0a^t_1)=0$,
  and the mass parameter equals to the moment of inertia of rigid sphere, ${\cal M}(_0a^t_1)={\cal J}=(2/5)MR^2$.
  It follows from Hamiltonian of normal vibrations, ${\cal H}=(1/2){\cal M}{\dot \alpha}^2+(1/2){\cal K}{\alpha}^2$,
  that in this dipole case a star sets in rigid-body rotation, rather than vibrations, about axis of its dipole magnetic moment; this feature of the model under consideration is quite similar to that of the neutron star model with homogeneous internal magnetic field. In Fig 3., we plot the frequency $\nu(_0a^t_\ell)$ and the period $P(_0a^t_\ell)=\nu^{-1}(_0a^t_\ell)$ of the Alfv\'en toroidal mode as functions of multipole degree computed in both homogeneous and nonhomogeneous neutron star models with indicated parameters.

  \begin{figure}[h]
\centering\
\includegraphics[width=8.0cm]{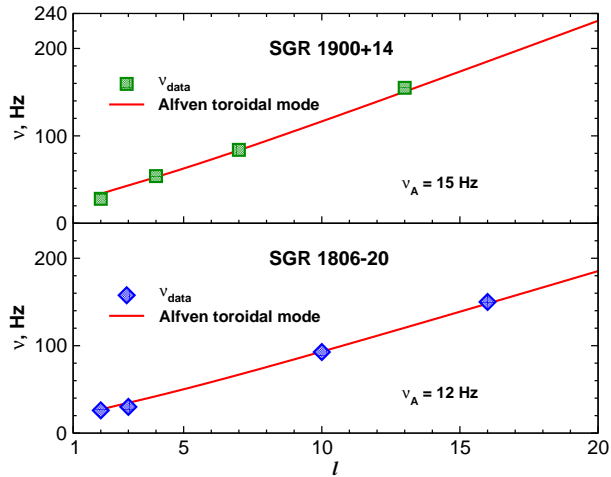}
\caption{\small
 Theoretical fit (lines) of data (symbols) on low-frequency QPOs during the flare of SGRs 1806-20 and 1900+14
 by the obtained spectral equation for the toroidal  Alfv\'en oscillations in Ferraro's poloidal field.}
\end{figure}

\section{QPOs in X-ray luminosity of flaring SGR 1806-20 and SGR 1900+14 from the viewpoint of considered
model}

As was mentioned, the one-parametric spectral formula (\ref{e1.8}), computed in the neutron star model with homogeneous internal and dipolar external magnetic field, does not reproduce general trends in data on QPOs frequencies whose numerical values for SGR 1806-20 are given by $\nu_{\rm data}=$18, 26, 30, 92, 150, 625, 1840 and for the SGR 1900+14 these are $\nu_{\rm data}=$28, 54, 84, 155 (Watts \& Strohmayer 2007).  It is tempting, therefore, to consider
these data from the view point of investigated model by identifying the observed QPOs with
overtones of spectral formula (\ref{e2.14}). The result of $\ell$-pole modal specification of the detected QPOs frequencies as overtones
of torsional Alfv\'en seismic vibrations in question is presented in Fig.4 and Fig.5. Specifically, for SGR 1900+14  we obtain: $\nu(_0a^t_2)=28$ Hz; $\nu(_0a^t_4)=53$; Hz
 $\nu(_0a^t_6)=84$ Hz; $\nu(_0t_{13})=155$ Hz, and for the SGR 1806-20 we get
 $\nu(_0a^t_2)=26$ Hz; $\nu(_0a^t_3)=30$, $\nu(_0a^t_{10})=92$ Hz; $\nu(_0a^t_{16})=155$ Hz; $\nu(_0a^t_{65})=625$ Hz and $\nu(_0a^t_{180})=1840$ Hz. While the lowest of detected oscillations, with $\nu=18$ Hz, cannot be specified
 in terms of considered seismic vibrations, it is clearly seen that the obtained spectrum correctly reflects
 general trends in the detected QPO frequencies. This suggests, if the detected QPOs are really produced by Lorentz-force-driven global nodeless torsional seismic vibrations about the dipole magnetic moment of magnetars, their internal magnetic fields should be of substantially nonhomogeneous configuration.

\begin{figure}[h]
\centering\
\includegraphics[width=12.0cm]{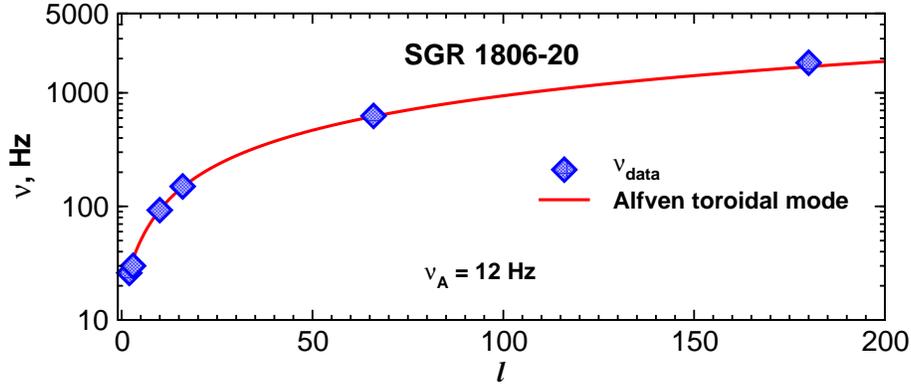}
\caption{\small
 Theoretical fit (lines) of data (symbols) on high-frequency QPOs frequencies
 during the flare of SGRs 1806-20 by the obtained spectral equation for the frequency of
 toroidal  Alfv\'en mode.}
\end{figure}

The result of $\ell$-pole modal specification of the detected QPOs frequencies as overtones
of torsional Alfv\'en seismic vibrations in question is presented in Fig.4 and Fig.5. Specifically, for SGR 1900+14  we obtain: $\nu(_0a^t_2)=28$ Hz; $\nu(_0a^t_4)=53$; Hz
 $\nu(_0a^t_6)=84$ Hz; $\nu(_0t_{13})=155$ Hz, and for the SGR 1806-20 we get
 $\nu(_0a^t_2)=26$ Hz; $\nu(_0a^t_3)=30$, $\nu(_0a^t_{10})=92$ Hz; $\nu(_0a^t_{16})=155$ Hz; $\nu(_0a^t_{65})=625$ Hz and $\nu(_0a^t_{180})=1840$ Hz. While the lowest of detected oscillations, with $\nu=18$ Hz, cannot be specified
 in terms of considered seismic vibrations, it is clearly seen that the obtained spectrum correctly reflects
 general trends in the detected QPO frequencies. This suggests, if the detected QPOs are really produced by Lorentz-force-driven global nodeless torsional seismic vibrations about the dipole magnetic moment of magnetars, their internal magnetic fields should be of substantially nonhomogeneous configuration.

\section{Concluding remarks}

 Any attempt to predict the behavior of Alfv\'en vibrational modes in pulsars and magnetars,
 presuming of course that the star material possesses properties of perfectly conducting continuous medium
 pervaded by magnetic fields, is beset with uncertainties regarding geometrical configuration of fossil internal magnetic field. It seems, therefore, that progress can be best made by studying these modes
 within the framework of comprehensive models. Among these are the models with homogeneous and nonhomogeneous axisymmetric poloidal magnetic fields considered long ago in works of Chandrasekhar and Fermi (1953) and by Ferraro (1954), respectively, to show that such fields
 have the same effect as rigid rotation, that is, tend to produce a flattening of the star shape along the magnetic field axis. Following this line of argument and continuing investigations reported in (Bastrukov et al 2009a),
 we have computed here the frequency spectrum of axisymmetric torsional nodeless vibrations, in the neutron star model with Ferraro's form of nonhomogeneous poloidal magnetic field which is presented in Fig.3 in juxtaposition
 in a neutron star model with homogeneous internal field.

 The practical usefulness of the obtained one-parametric spectral formula has been demonstrated by its application
 to $\ell$-pole identification of QPOs frequencies during the X-ray giant outbursts of SGR 1900+14 and SGR 1806-20.
 The result of our analysis, summarized in Fig.4 and Fig.5, shows that the model adequately regains the overall trends in the detected QPOs frequencies and, thus, supports theoretical interpretation of these QPOs, advanced in works reporting this discovery (Israel et al 2005, Watts \& Strohmayer 2006), as owing their origin to quake-induced torsional seismic vibrations of underlying magnetar. Together with this, in (Bastrukov et al 2009b) it has been shown that the same data on the QPO frequencies can be consistently interpreted from the view point of two-component, core-crust, model of quaking neutron star (Franco et al. 2000) with homogeneous internal magnetic field as being produced by axisymmetric differentially rotational Alfv\'en nodeless oscillations of crustal solid-state plasma about axis of magnetic field frozen in the immobile core. With all these in mind, we conclude that it is
 the Lorentz force of magnetic field stresses plays decisive part in post-quake vibrational relaxation of above magnetars and that the toroidal fields of quake-induced material displacements are of substantially nodeless character.

 Authors are grateful to Dr. Dima Podgainy for helpful assistance. This work has been supported by NSC of
 Taiwan, grant numbers  NSC-098-2811-M-007-009 and NSC-96-2628-M-007-012-MY3.

\appendix

\section{Torsional stiffness of global torisonal Alfv\'en nodeless oscillations of a neutron star
about axis of Ferraro's nonhomogeneous poloidal magnetic field}

 In computing stiffness of torsional Alfv\'en oscillations
  \begin{eqnarray}
\nonumber
 && {\cal K}_m=\int \tau_{ik}({\bf r})\,a_{ik}({\bf r})\,d{\cal V}, \quad  a_{ik}({\bf r})=\frac{1}{2}[\nabla_i\,a_k({\bf r})+\nabla_k\,a_i({\bf r})],\\
\nonumber
 && \tau_{ik}({\bf r})=\frac{1}{4\pi}[B_i({\bf r})\,b_k({\bf r})+B_k\,({\bf r})b_i({\bf r})],\quad b_i({\bf r})=(B_k ({\bf r})\nabla_k)a_i({\bf r})-(a_k({\bf r})\nabla_k) B_i({\bf r})
  \end{eqnarray}
it is convenient to represent strain tensor
 \begin{eqnarray}
 \nonumber
 a_{ik}=\frac{1}{2}(\nabla_i a_k + \nabla_k a_i)
 \end{eqnarray}
 in spherical polar coordinates with use of the angle variable ${\zeta}=\hbox{cos}\,{\theta}$.
 In terms of this variable, the components of these tensor are
 \begin{eqnarray}
 \nonumber
 &&a_{rr}=\frac{\partial a_r}{\partial r},\quad
 \quad\quad
 a_{\theta\theta}=-\frac{(1-\zeta^2)^{1/2}}{r}
 \frac{\partial a_r}{\partial \zeta}+\frac{a_r}{r},\\
 \nonumber
 && a_{\phi\phi}=\frac{1}{r}\frac{1}{(1-\zeta^2)^{1/2}}
 \frac{\partial{a_\phi}}{\partial{\phi}}+\frac{a_r}{r}+
 \frac{\zeta}{(1-\zeta^2)^{1/2}}\frac{a_\theta}{r},
 \\
 \nonumber
 &&a_{r\theta}=\frac{1}{2}\left[-\frac{(1-\zeta^2)^{1/2}}{r}
 \frac{\partial a_r}{\partial \zeta}-\frac{a_\theta}{r}+
 \frac{\partial a_\theta}{\partial r}\right],
 \\
 \nonumber
 &&a_{r\phi}=\frac{1}{2}\left[\frac{1}{r}\frac{1}{(1-\zeta^2)^{1/2}}
 \frac{\partial a_r}{\partial \phi}-\frac{a_\phi}{r}+
 \frac{\partial a_\phi}{\partial r}\right],
 \\
 \nonumber
 &&a_{\theta\phi}=\frac{1}{2}\left[\frac{1}{r}\frac{1}{(1-\zeta^2)^{1/2}}
 \frac{\partial a_\theta}{\partial \phi}-
 \frac{\zeta}{(1-\zeta^2)^{1/2}}\frac{a_\phi}{r}-
 \frac{(1-\zeta^2)^{1/2}}{r}\frac{\partial a_\phi}{\partial \zeta}\right].
 \end{eqnarray}
 In the torsional mode of nodeless vibrations the field of instantaneous displacements
 has solely one non-zero $\phi-th$ component
 \begin{eqnarray}
 \nonumber
 a_r=0\quad a_\theta=0\quad a_{\phi}=A_t\,r^\ell(1-\zeta^2)^{1/2}P'_\ell(\zeta),\quad
 P'_\ell(\zeta)=\frac{dP_\ell(\zeta)}{d\zeta}.
 \end{eqnarray}
 In this case we have only two non-zero components of the strain tensor
 \begin{eqnarray}
 \nonumber
&& a_{rr}=a_{\theta\theta}=a_{\phi\phi}=a_{r\theta}=0,\\
 \nonumber
 && a_{r\phi}=\frac{A_t}{2}r^{\ell-1}
   (\ell-1)(1-\zeta^2)^{1/2}P'_\ell,\quad a_{\theta\phi}=-\frac{A_t}{2}r^{\ell-1}
   \left[2\zeta P'_\ell-\ell(\ell+1)P_\ell(\zeta)\right].
 \end{eqnarray}
In spherical polar coordinates the components of vector field $b_i({\bf r})=(B_k \nabla_k)a_i-(a_k\nabla_k) B_i$
are given by
 \begin{eqnarray}
 \nonumber
 b_r&=&\left[B_r\frac{\partial }{\partial r}-\frac{B_\theta}{r}(1-\zeta^2)^{1/2}\frac{\partial }{\partial \zeta}+
 \frac{B_\phi}{r}(1-\zeta^2)^{-1/2}\frac{\partial }{\partial \phi}\right]\,a_r-
 \frac{B_\theta\,a_\theta+B_\phi\,a_\phi}{r}\\
 \nonumber
 &-&
 \left[a_r\frac{\partial }{\partial r}-
 \frac{a_\theta}{r}(1-\zeta^2)^{1/2}\frac{\partial }{\partial \zeta}+
 \frac{a_\phi}{r}(1-\zeta^2)^{-1/2}\frac{\partial }{\partial \phi}\right]\,B_r+
 \frac{a_\theta\,B_\theta+a_\phi\,B_\phi}{r},
 \\[0.3cm]
 \nonumber
 b_\theta&=&\left[B_r\frac{\partial }{\partial r}-\frac{B_\theta}{r}(1-\zeta^2)^{1/2}\frac{\partial }{\partial \zeta}+
 \frac{B_\phi}{r}(1-\zeta^2)^{-1/2}\frac{\partial }{\partial \phi}\right] \,a_\theta
 +\frac{B_\theta\,a_r-B_\phi\,a_\phi\,\zeta(1-\zeta^2)^{-1/2}}{r}\\
 \nonumber
 &-&\left[a_r\frac{\partial }{\partial r}-
 \frac{a_\theta}{r}(1-\zeta^2)^{1/2}\frac{\partial }{\partial \zeta}+
 \frac{a_\phi}{r}(1-\zeta^2)^{-1/2}\frac{\partial }{\partial \phi}\right]
 \,B_\theta
 -\frac{a_\theta\,B_r-a_\phi\,B_\phi\,\zeta(1-\zeta^2)^{-1/2}}{r},
 \\[0.3cm]
 \nonumber
 b_\phi&=&\left[B_r\frac{\partial }{\partial r}-\frac{B_\theta}{r}(1-\zeta^2)^{1/2}\frac{\partial }{\partial \zeta}+
 \frac{B_\phi}{r}(1-\zeta^2)^{-1/2}\frac{\partial }{\partial \phi}\right]\,
 a_\phi+
 \frac{B_\phi\,a_r+B_\phi\,a_\theta\,\zeta(1-\zeta^2)^{-1/2}}{r}\\
 \nonumber
 &-&\left[a_r\frac{\partial }{\partial r}-
 \frac{a_\theta}{r}(1-\zeta^2)^{1/2}\frac{\partial }{\partial \zeta}+
 \frac{a_\phi}{r}(1-\zeta^2)^{-1/2}\frac{\partial }{\partial \phi}\right]\,
 B_\phi-\frac{a_\phi\,B_r+a_\phi\,B_\theta\,\zeta(1-\zeta^2)^{-1/2}}{r}.
 \end{eqnarray}
Taking into account that Ferraro's field has only two non-zero components which can be conveniently represented
in the form
\begin{eqnarray}
 \nonumber
B_r=\frac{2f}{r^2}\,\zeta,\quad B_\theta=-\frac{(1-\zeta)^{1/2}}{r}\,f',\quad B_\phi=0,\quad
f=\frac{B}{4R^2}\,r^2(3r^2-5R^2),\quad f'=\frac{df}{dr}
\end{eqnarray}
for the components of $b_i$ we obtain
 \begin{eqnarray}
 \nonumber
 b_r=0,\quad b_\theta=0, \quad b_\phi=B_r\frac{\partial a_\phi}{\partial r}-
 \frac{B_\theta}{r}(1-\zeta^2)^{1/2}\frac{\partial a_\phi}{\partial \zeta}
 -\frac{a_\phi\,B_r}{r}-\frac{a_\phi\,B_\theta\,\zeta(1-\zeta^2)^{-1/2}}{r}
 \end{eqnarray}
 where
  \begin{eqnarray}
 \nonumber
  \frac{\partial a_\phi}{\partial r}&=&A_t\,\ell\,r^{\ell-1}(1-\zeta^2)^{1/2}P'_\ell,\quad\quad
  \frac{\partial a_\phi}{\partial \zeta}=A_t\,r^\ell\,\left(1-\zeta^2 \right)^{-1/2}\left[\zeta\,P'_\ell-\ell\left(\ell+1 \right)  \right].
  \end{eqnarray}
 The integrand of ${\cal K}_m$ reads
 $$\tau_{ik}a_{ik}=2\left( \tau_{r\phi}a_{r\phi}+\tau_{\theta\phi}a_{\theta\phi}\right)$$
 so that relevant to computation of ${\cal K}_m$ components of tensor $\tau_{ik}=(1/4\pi)[B_i\,b_k+B_k\,b_i]$ are given by
\begin{eqnarray}
 \nonumber
 \tau_{r\phi}&=&\frac{1}{4\pi}\left[B_rB_r\frac{\partial a_\phi}{\partial r}-
 \frac{B_rB_\theta}{r}(1-\zeta^2)^{1/2}\frac{\partial a_\phi}{\partial \zeta}\,
 -\frac{a_\phi\,B_rB_r}{r}-\frac{a_\phi\,B_rB_\theta\,\zeta(1-\zeta^2)^{-1/2}}{r}\right] \\
 \nonumber
 &=&\frac{A_t}{4\pi}\left\{4(\ell-1)f^2 r^{\ell-5} \zeta^2(1-\zeta^2)^{1/2}P'_\ell + 2f f'r^{\ell-4}\zeta(1-\zeta^2)^{1/2}[2\zeta P'_\ell-\ell(\ell+1)P_\ell]\right\},\\
 \nonumber
 \tau_{\theta\phi}&=&\frac{1}{4\pi}\left[B_\theta B_r\frac{\partial a_\phi}{\partial r}-
 \frac{B_\theta B_\theta}{r}(1-\zeta^2)^{1/2}\frac{\partial a_\phi}{\partial \zeta}\,
 -\frac{a_\phi\,B_\theta B_r}{r}-\frac{a_\phi\,B_\theta\, B_\theta\,\zeta(1-\zeta^2)^{-1/2}}{r}\right]\\
  \nonumber
 &=&\frac{A_t}{4\pi}\left[-2r^{\ell-4}\left(\ell-1 \right)f f'\zeta\left(1-\zeta^2 \right)P'_\ell -r^{\ell-3}f'^2\left(1-\zeta^2\right)\left[2\zeta P'_\ell-\ell\left( \ell+1\right)P_\ell  \right]\right].
 \end{eqnarray}
The integral for stiffness can be conveniently represented in the form
\begin{eqnarray}
 \nonumber
 {\cal K}_m&=&2\int [ \tau_{r\phi}\,a_{r\phi}+\tau_{\theta\phi}\,a_{\theta\phi}]d{\cal V}=\frac{A_t^2}{2}\left\{4(\ell-1)^2R_{ff}I_1+4(\ell-1)R_{ff'}[2I_1-\ell(\ell+1)I_2] \right . \\
 \nonumber
 &+&
 \left.
 R_{f'f'}[4I_1-4\ell(\ell+1)I_2+\ell^2(\ell+1)^2I_3]\right\}
 \end{eqnarray}
The integrals $I_i$ are computed with aid of standard recurrence relations between Legendre polynomials (e.g. Abramowitz \& Stegan 1964) which yield
\begin{eqnarray}
\nonumber
&&I_1=\int\limits_{-1}^{1}\zeta^2(1-\zeta^2) (P'_\ell)^2 d\zeta
     =\frac{2\,\ell\,(\ell+1)(2\ell^2+2\ell-3)}{(4\ell^2-1)(2\ell+3)},\\ \nonumber
&&I_2=\int\limits_{-1}^{1}\zeta(1-\zeta^2)P_\ell\,P'_\ell d\zeta
     =\frac{2\,\ell\,(\ell+1)}{(4\ell^2-1)(2\ell+3)},\\ \nonumber
&&I_3=\int\limits_{-1}^{1}(1-\zeta^2)P_\ell^2 d\zeta=\frac{4\,(\ell^2+\ell-1)}{(4\ell^2-1)(2\ell+3)},\quad I_4=\int\limits_{-1}^{1}\zeta^2(1-\zeta^2) P_\ell P'_\ell d\zeta
     =\,0.
\end{eqnarray}
For integrals with function $f=(B/4R^2)[r^2(3r^2-5R^2)]$ we obtain
\begin{eqnarray}
\nonumber
&& R_{ff}=\int_0^R f^2(r) r^{2\ell-4} dr= \frac{B^2R^{2\ell+1}}{16}R_1,\quad R_1=\left[\frac{25}{2\ell+1}-\frac{30}{2\ell+3}+
\frac{9}{2\ell+5}\right],\\
\nonumber
&&  R_{ff'}=\int_0^R f\left(\frac{df}{dr}\right) r^{2\ell-3} dr=\frac{B^2R^{2\ell+1}}{16}R_2,\quad
R_2=\left[\frac{50}{2\ell+1}
-\frac{90}{2\ell+3}+\frac{36}{2\ell+5}\right],\\
\nonumber
&& R_{f'f'}= \int_0^R \left(\frac{df}{dr}\right)^2 r^{2\ell-2} dr=\frac{B^2R^{2\ell+1}}{16}R_3,\quad R_3=\left[\frac{100}{2\ell+1}-\frac{240}{2\ell+3}+
\frac{144}{2\ell+5}\right].
\end{eqnarray}
The resultant expression for the stiffness reads
$${\cal K}_m=B^2\,A_t^2\,R^{2\ell+1}\, k_\ell$$
where
\begin{eqnarray}
 \nonumber
 k_\ell&=&\frac{1}{32}\left\{4(\ell-1)^2R_1\,I_1+4(\ell-1)R_2[2I_1-\ell(\ell+1)I_2]+  R_3[4I_1-4\ell(\ell+1)I_2+\ell^2(\ell+1)^2I_3]\right\}\\ \nonumber
 &=&\frac{\ell \left(\ell^2-1 \right)   }
 {2\left(4\ell^2-1 \right)\left(2\ell+3 \right) \left(2\ell+5 \right) }\left(5\ell^3+7\ell^2+59\ell+84 \right).
 \end{eqnarray}

\end{document}